\documentclass[preprint,prl]{revtex4}

\usepackage{graphicx}

\begin{document}

\title{Spontaneous natural selection in a model for spatially distributed interacting populations}

\author{Monica F. B. Moreira}
\affiliation{Departamento de Ci\^encias Exatas, Universidade Federal de Lavras, 37200-000 Lavras, MG,
Brazil}
\author{M\'arcio P. Dantas}
\affiliation{Departamento de Ci\^encias Exatas, Universidade Federal de Lavras, 37200-000 Lavras, MG,
Brazil}
\author{A. T. Costa Jr.}
\affiliation{Departamento de Ci\^encias Exatas, Universidade Federal de Lavras, 37200-000 Lavras, MG,
Brazil}

\begin{abstract}

We present an individual-based model for two interacting populations
diffusing on lattices in which a strong natural selection develops
spontaneously. The models combine traditional local predator-prey
dynamics with random walks. Individual's mobility is considered as
an inherited trait. Small variations upon inheritance, mimicking
mutations, provide variability on which natural selection may
act. Although the dynamic rules defining the models do not explicitly
favor any mobility values, we found that the average mobility of both
populations tend to be maximized in various situations. In some
situations there is evidence of polymorphism, indicated by an adaptive 
landscape with many local maxima. We provide evidence relating 
selective pressure for high mobility with pattern formation.

\end{abstract} 

\maketitle

The possibility of modeling certain aspects of biological evolution fascinates 
physicists and mathematicians. One of the most attractive aspects of evolution
is its apparent self-organization. Properties that emerge spontaneously from
the system's dynamics, like evolutionary stable strategies (ESS), for instance, are
ubiquitous in nature. One important goal of population biology is to understand
the relationship between system's dynamics and evolutionary processes.
To this end mathematical modeling may be one of the most valuable tools.
The possibility of building models that isolate the fundamental features
of a system responsible for some specific behavior is specially attractive.
In the study of evolutionary processes one important nuisance is the time
scale in which evolution occurs. Mathematical modeling is specially important
in this respect, since it allows one to test many different evolutionary histories
in relatively short times, a possibility obviously not available experimentally.

Co-evolution is one of the most fascinating subjects in population
biology. It is also fertile ground for mathematical modeling. The
combination of non-linear dynamics with evolutionary processes has a huge
potential for generating complex behaviors. One of these is speciation.
There is much interest in processes leading to sympatric speciation,
specially on the light of recent experimental evidence \cite{via_2001,turelli_2001,friesen_2004}.
It is customary, when modeling sympatric speciation, to assume adaptive 
landscapes with multiple maxima, which forcibly generates populations
with as many kinds of individuals as there are local fitness maxima \cite{karen}.
This situation is known in biology as a polymorphic population, and is the first
step in speciation \cite{mayr_EvolDiv,ridley_evolution}. One intriguing possibility, 
however, is that polymorphic populations arise as a result of the system's dynamics 
itself, without need to impose by hand a complex adaptive landscape. 
In this letter we show that in a model for spatially distributed populations 
coupled by predator-prey dynamics natural selection arises spontaneously 
and, in some cases, the adaptive landscape has more than one maximum, which
results in the appearance of polymorphic populations. In other words, spatial distribution
coupled to predator-prey dynamics generates dynamically a non-flat 
adaptive landscape, that may even be complex to the point of exhibiting 
multiple maxima. This indicates that the complexity of an adaptive landscape
need not be consequence of complex environments, but may be the result
of the system dynamics alone.

Another interesting aspect of our results is pattern formation. The
``environment'' we consider is spatially homogeneous. However,
in many situations the populations distribute themselves inhomogeneously.
This is also a dynamically generated effect that occurs even in the absence
of an evolutionary mechanism \cite{murray}. As we will show, it is possible to correlate 
pattern formation and emergence of natural selection in the model we are studying.
As a matter of fact, there seems to be a complicated interplay between
the mechanisms primarily responsible for pattern formation and 
the evolutionary mechanism, evidenced by the fact that, in some situations,
the action of the evolutionary mechanism may change pattern formation
drastically.

\textit{The Model:} We model the habitat as a one-dimensional lattice
with $N$ sites, and periodic boundary conditions. Individuals move
through the lattice in a random walk, characterized by a parameter
we call \textsl{mobility}, denoted by $M$. It is the probability that
an individual moves to one of the nearest-neighbor sites in a given 
time step. Thus, $M\in[0,1]$. Both populations obey probabilistic
(asexual) reproduction and death rules, inspired by the Lotka-Volterra equations.
Namely, an individual of the prey population at site $i$ and time step $t$
reproduces with probability $p_{repr}^{prey}(i,t)=\alpha(1-Y(i,t)/K)$, where
$Y(i,t)$ is the prey population at site $i$ and time step $t$, $\alpha$
is an intrinsic reproductive rate and $K$ is the support capacity of each
lattice site, which we assume to be the same at all lattice sites. 
Predators at site $i$ kill prays with probability proportional to the prey 
population at that site, and reproduce at a rate proportional to the number 
of preys killed. Predators also die at a constant rate $\mu$. The local dynamics just presented
are equivalent on average, for large populations, to the Lotka-Volterra equations with
a Verhulst term \cite{solange_2001}. The random walk is, on average, equivalent to
Fick diffusion; the combination of both mechanisms leads to
reaction-diffusion equations, which are well-known as mathematical models for
pattern formation \cite{murray}.

We chose the individual's mobility $M$ as its inherited trait. This choice may
seem somewhat arbitrary, but many real populations have evolutionary strategies
that are strongly tied to their ability of moving through its habitat \cite{doncaster_2003}. 
Thus, in our model, each individual inherits, at birth, its parent's mobility $M$
plus a small random variation $\delta M$ taken from
the interval $[-\sigma,\sigma]$ with uniform probability.
$\delta M$ plays the role of mutation and $\sigma$ is 
associated with the mutation rate. Thus variability is generated
in the population, and if there is any selective mechanism in
action we expect to see evolution. In our case, evolution would 
be characterized by a systematic displacement of the mean
population mobility. We should also be able to perceive the
effects of natural selection through observation of the 
population's mobility distribution. If there is selective pressure 
related to mobility, it is reasonable to expect a mobility distribution 
concentrated around some ``optimum'' mobility.

It is worth stressing that there is no explicit adaptive advantage
associated with any mobility value. Any selective pressure we happen
to find is a result solely of the system's dynamics.

\begin{figure}
\includegraphics[width=0.8\columnwidth,clip]{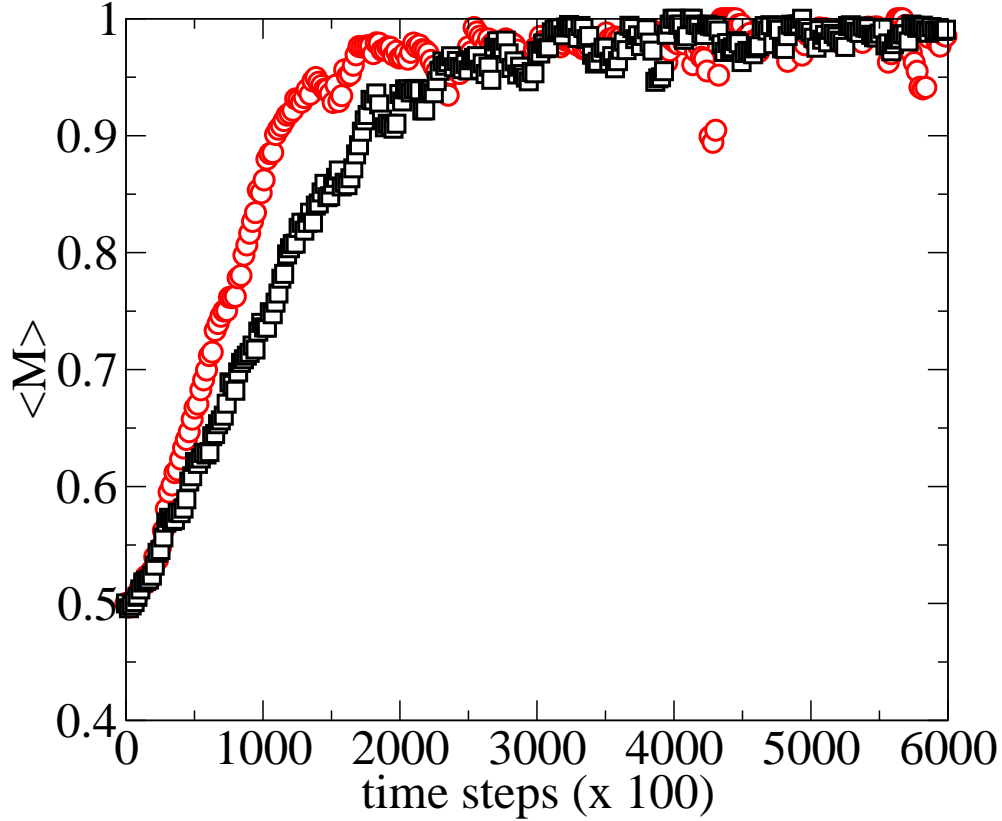}
\caption{Average mobility $\langle M\rangle$ of predator (squares) and 
prey (circles) populations, as a function of time, for the first set of parameters. 
At the beginning ($t=0$) all individuals in both populations have the 
same mobility ($M=0.5$).}
\label{avMob_1}
\end{figure}

\begin{figure}
\includegraphics[width=0.8\columnwidth,clip]{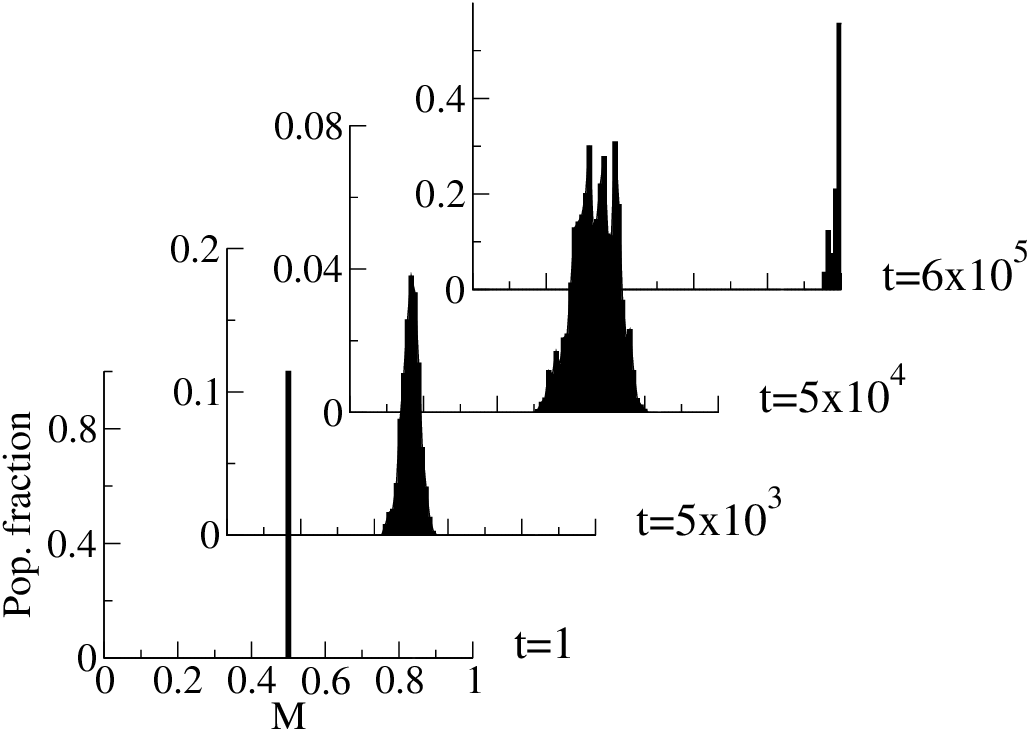}
\caption{Mobility distributions of prey population (predator's has similar behavior)
for selected time steps, starting with concentrated distributions at time $t=0$.
The set of parameters is the same as in figure~\ref{avMob_1}.}
\label{MobDistrib_1}
\end{figure}

\textit{Results:} We start by showing how the average mobilities $\langle M\rangle$
of both populations (predator and prey) evolve in time, Fig.~\ref{avMob_1},
for a chosen set of parameters, to which we will refer as the first set. 
It clearly increases up to a saturation value that is very close to the
maximum probability of 1. The high rate at which $\langle M\rangle$ increases
suggests a strong selective pressure towards high mobilities. By analyzing the 
corresponding mobility distributions, Fig.~\ref{MobDistrib_1}, we confirm that
selection is indeed acting upon both populations. Starting
from an extremely concentrated distribution (all the individuals with
the same mobility at $t=0$), there appears some variability due to
mutations, but this variability is kept small at all times. If there were
no selection mechanism at work mutations would steadily increase 
the width of the mobility distribution towards uniformity. 

\begin{figure}
\includegraphics[width=0.8\columnwidth,clip]{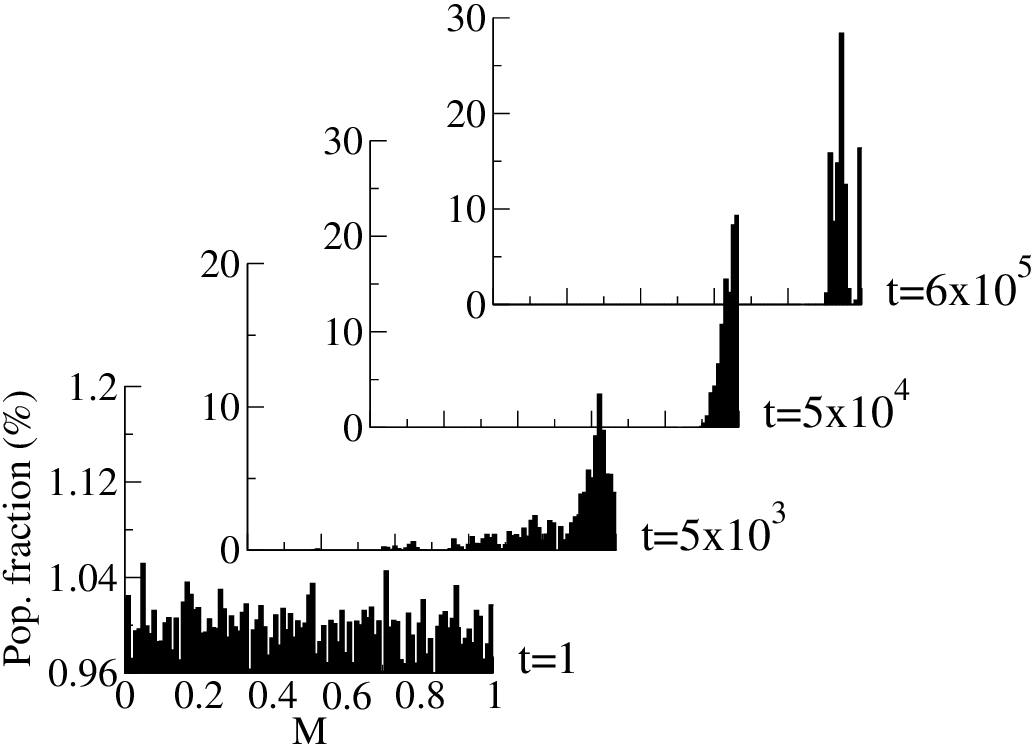}
\caption{Mobility distributions of  prey population (predator's has similar behavior)
for selected time steps, starting with uniform distributions at time $t=0$. The set 
of parameters is the same as in figure~\ref{avMob_1}.}
\label{MobDistrib_1a}
\end{figure}

The fact that we chose initial distributions for mobility concentrated
around a single mobility value and that this distribution moved steadily
to high $\langle M\rangle$ values suggests that the adaptive landscape is
a simple one, with a single maximum near $\langle M\rangle=1$. It is worth
inquiring further into the structure of the adaptive landscape by choosing
a different initial condition: individuals are assigned random mobility values 
taken from an uniform distribution over the interval $[0,1]$. In this way, every region
of the adaptive landscape is accessible to at least a few individuals; if a region
contains an adaptation maximum it is very probable that the individuals in that
region will form a significant fraction of the final population. On the other hand,
individuals that happen to be around a minimum will most certainly disappear
from the population relatively quickly.
As Fig.~\ref{MobDistrib_1a} shows, the distribution is rapidly narrowed,
once again evidencing a strong selective pressure in favor of high mobilities.
Although we expected, from the previous results, the adaptive landscape to
be simple, exhibiting a single maximum very near $\langle M\rangle=1$, we now 
see a hint of a more complex structure, with two maxima. Nevertheless both
maxima correspond to very similar values of mobility, and may be artifacts
of statistical fluctuations in finite populations.

\begin{figure}
\includegraphics[width=0.8\columnwidth,clip]{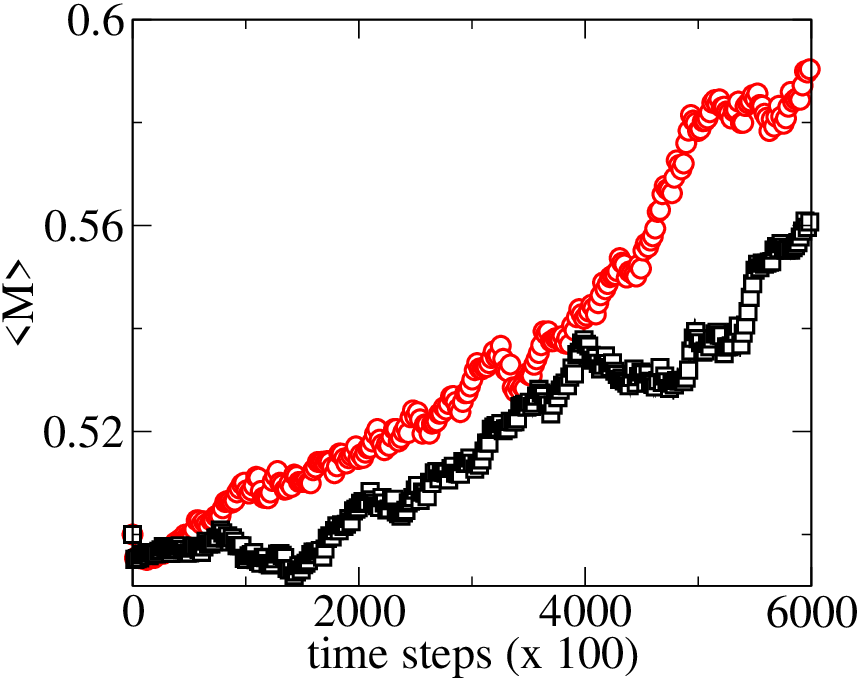}
\caption{Average mobility $\langle M\rangle$ of predator (squares) and prey (circles) 
populations, as a function of time, for the second set of parameters. At the 
beginning ($t=0$) all individuals in both populations have the same mobility ($M=0.5$).}
\label{avMob_2}
\end{figure}

\begin{figure}
\includegraphics[width=0.8\columnwidth]{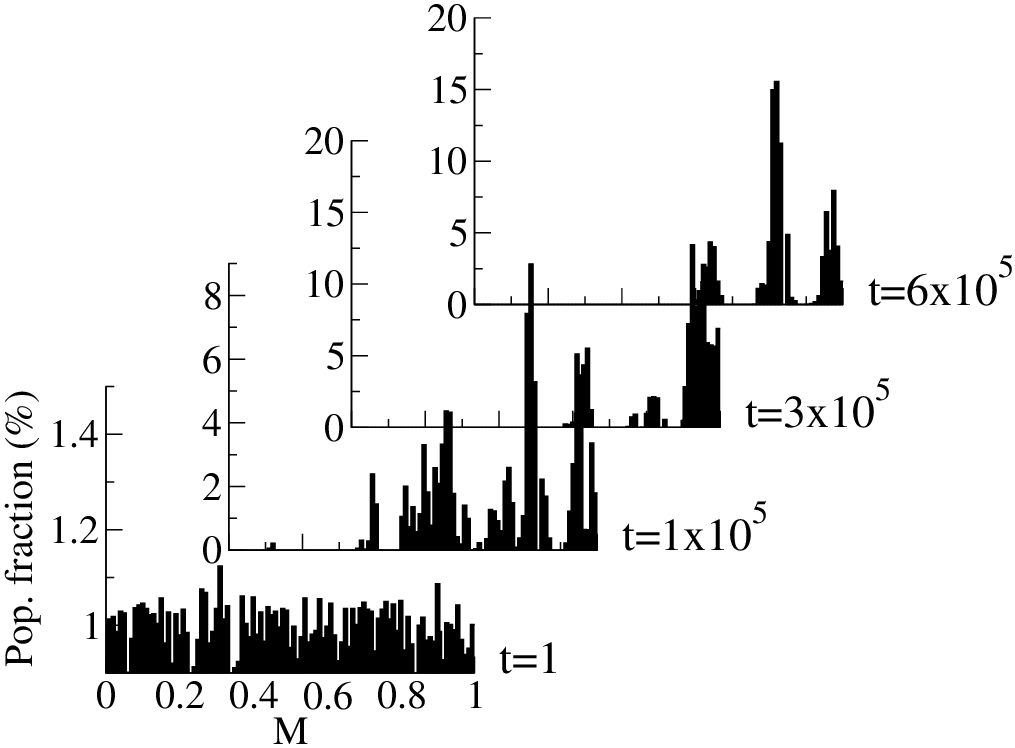}
\includegraphics[width=0.8\columnwidth,clip]{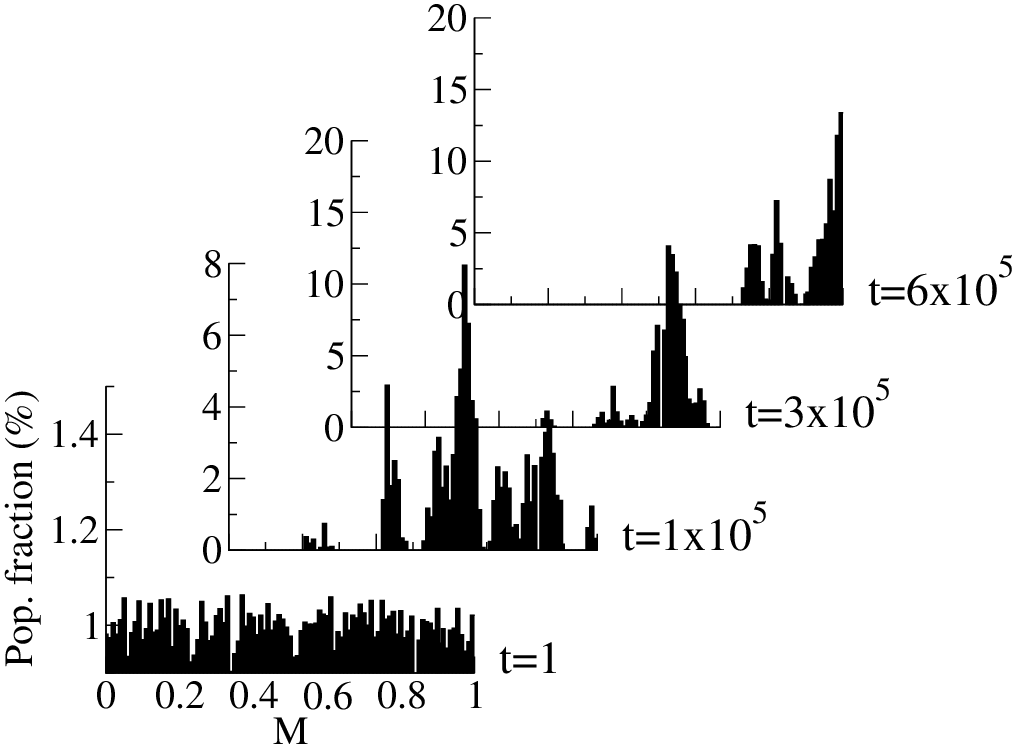}
\caption{Mobility distributions of predator (top panel) and prey (bottom panel) 
populations for selected time steps, starting with uniform distributions at time $t=0$.
The second set of parameters has been used. Notice that the asymptotic distributions
display at least three distinct maxima, indicating a polymorphic population.}
\label{MobDistrib_2a}
\end{figure}

If we change the set of parameters regulating local dynamics,
call it the second set, different evolutionary regimes may be observed, as shown in
Fig.~\ref{avMob_2}.  There are outstanding differences between these
results and those presented on Fig.~\ref{avMob_1}: the saturation
value reached by the average mobility is much smaller in this second
case, as is its rate of change. This suggests that the selective
pressure is less intense in this case. However, the smaller saturation
value may have a diverse meaning: since we started with a concentrated
mobility distribution and the mutation rate is relatively small, if
the adaptive landscape has more than one maximum the population may
be ``trapped'' in a local maximum, much like a dynamical system may
become ``trapped'' in a metastable state. To settle the question we
again resort to an initial distribution of mobilities that is
uniform. Here we see that this situation is a bit more complex: the
adaptive landscape displays multiple maxima indeed, as shown in
Fig.~\ref{MobDistrib_2a}. The lower saturation value we observed in
Fig.~\ref{avMob_2} thus is a result of the population being trapped by
the first maximum it encounters. Since the mutation rate $\sigma$ is
relatively small, there is not enough variability for the population
to overcome the adaptive valley before reaching for the next peak.
We also have results showing that, if the mutation rate is increased
slightly, the excess variability thus generated is enough for the
transposition of the adaptive valley

\begin{figure}
\includegraphics[width=0.8\columnwidth,clip]{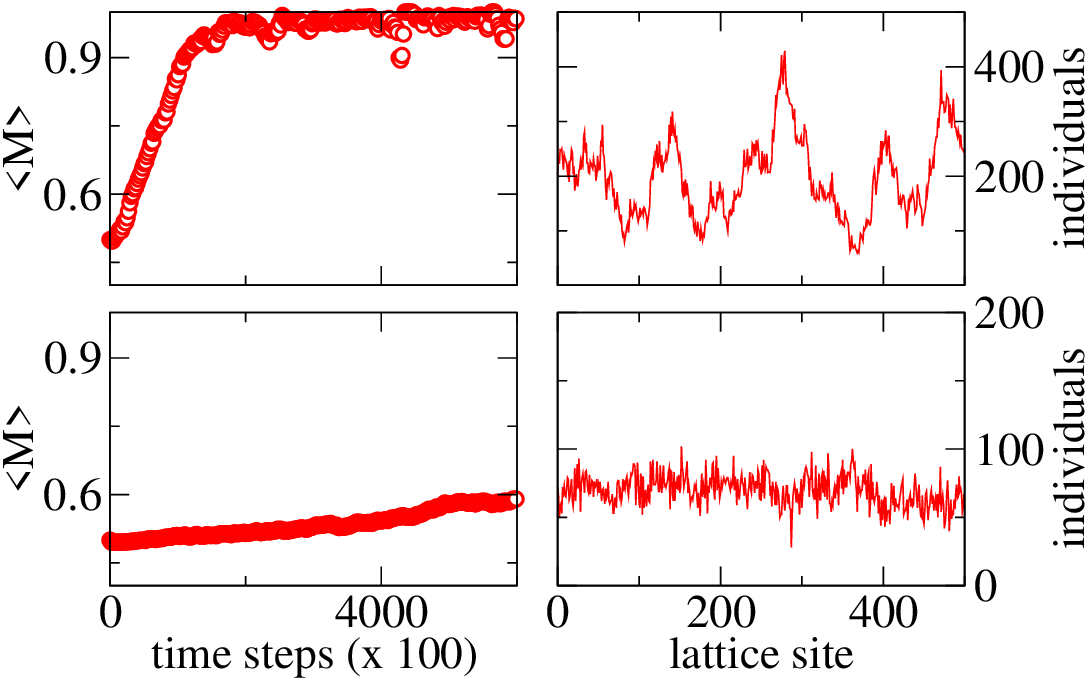}
\caption{Time evolution of $\langle M\rangle$ (left panel) for preys  and
corresponding asymptotic spatial distribution (right panel) of prey populations,
showing evidence of a correlation between heterogeneity in spatial distribution
and strength of natural selection.}
\label{pattern_selection}
\end{figure}

\begin{figure}
\includegraphics[width=0.8\columnwidth,clip]{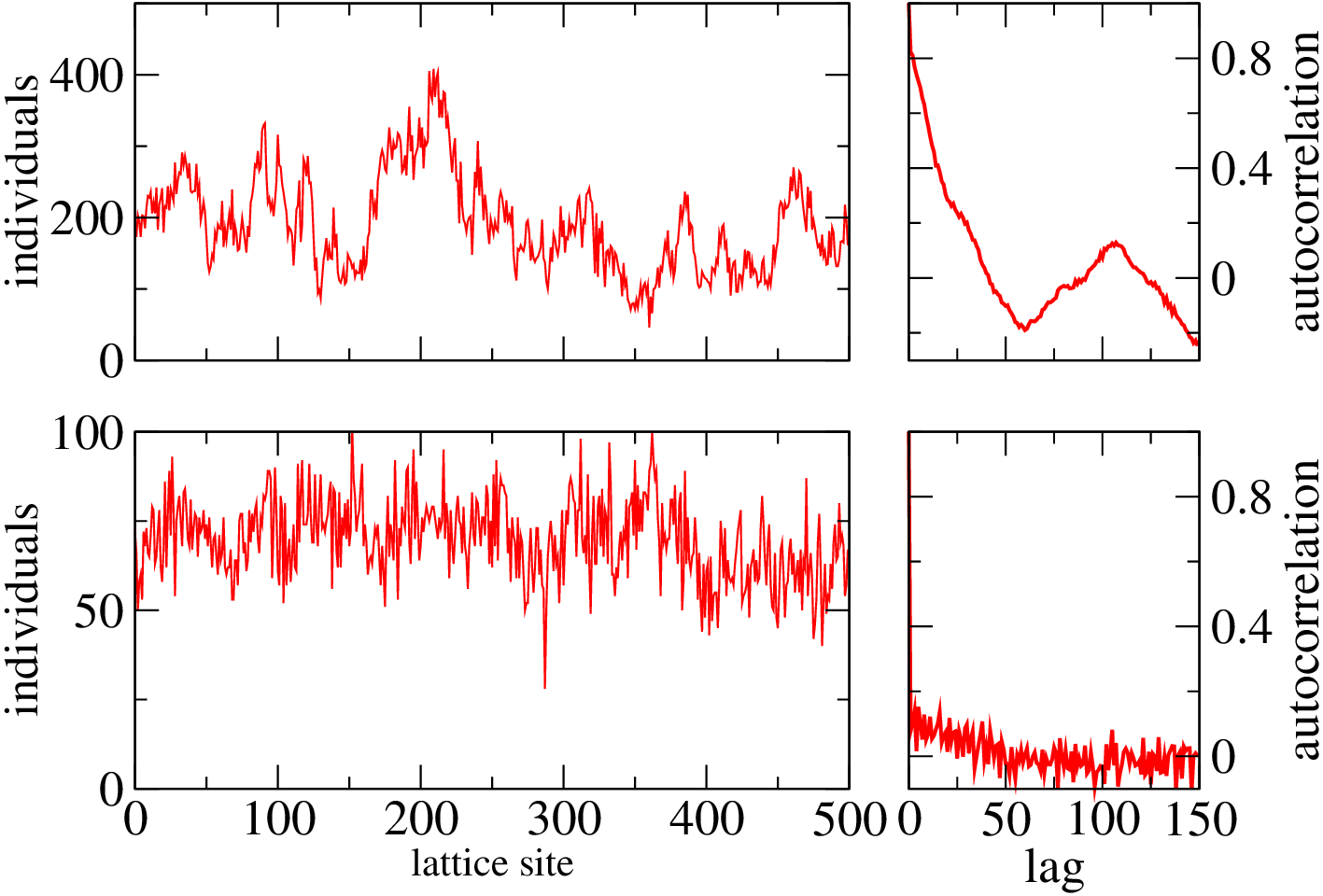}
\caption{Asymptotic spatial distribution (left panel) of prey in the absence (top)
and in the presence (bottom) of evolutionary mechanism. In the left panel,
the spatial autocorrelation function in both cases show that the evolutionary mechanism
changed qualitatively the spatial pattern, destroying the long-range order observed in the
absence of evolution.} 
\label{evol_changes_pattern}
\end{figure}

It is worth noticing that, besides its interesting evolutionary behavior, this
model also exhibits pattern formation, that is, even though the dynamical rules
are the same for the entire lattice, the resulting dynamics generates inhomogeneous
population distributions. This is not, however, a consequence of the evolutionary
mechanism but of the reaction-diffusion structure of the population dynamics.
Nevertheless, the evolutionary mechanism affects significantly the features of
the asymptotic patterns, and vice-versa. That this is so may be noticed 
from evidences of correlation between pattern formation and selective pressure.
These can be seen in Fig.~\ref{pattern_selection} where one sees that the selective 
pressure is stronger when the populations are distributed inhomogeneously in the environment.
Reasoning na\"ively one could argue that this result is somehow expected, since 
there is no apparent advantage in being more or less mobile in a homogeneous
environment. However, as stated above, pattern formation is a consequence of dynamics, 
as much as the selective pressure in our model and, in some situations, introduction
of the evolutionary mechanism changes qualitatively the population distribution,
as can be seen in Fig.~\ref{evol_changes_pattern}.
Thus, the relation between pattern formation and selective pressure is by no means 
trivial, and deserves further investigation.

To summarize, we presented a model for two spatially
distributed, co-evolving populations interacting via predator-prey
local dynamics, in which natural selection appears \textsl{spontaneously}
as a result of the system's dynamics. We also showed that in some cases
a complex adaptive landscape is generated dynamically, which may lead
to the appearance of polymorphic populations. This state is the departure point
for an important evolutionary process known as sympatric speciation. Incidentally
we showed that this model displays pattern formation, which may be
relevant to explain spontaneously fragmented habitats. We also showed evidence
of a correlation between pattern formation and selective pressure, that needs to
be better understood.

The authors acknowledge enlightening discussions with Thadeu J. P. Penna, 
Karen Luz-Burgoa, Jos\'e Nogales, J. N. C. Louzada and Iraziet C. Charret.
A.T.C. gratefully acknowledges Luciana J. Costa for many fruitful discussions 
and a critical reading of the manuscript. A.T.C. and M.P.D. acknowledge partial 
financial support from CNPq (Brazil). M.F.B.M. acknowledges partial financial 
support from CAPES (Brazil). The computational facilities in which part of this 
work was done were provided by FINEP (Brazil).


\end{document}